\documentclass[12pt]{article}
\usepackage{axodraw}

\catcode`@=12
\begin{document}
\thispagestyle{empty}
\begin{flushright} 
UCRHEP-T361\\ 
August 2003\
\end{flushright}
\vspace{1.0in}
\begin{center}
{\LARGE	\bf Neutrino Mass and the $SU(2)_R$ Breaking Scale\\}
\vspace{1.5in}
{\bf Ernest Ma\\}
\vspace{0.2in}
{\sl Physics Department, University of California, Riverside, 
California 92521\\}
\vspace{2.0in}
\end{center}
\begin{abstract}\
The neutrino sector in a left-right extension of the Standard Model depends 
on how $SU(2)_R$ is broken.  I list all possible scenarios, including the 
ones where the Majorana $\nu_R$ mass is naturally much smaller than the 
$SU(2)_R$ breaking scale, which is desirable for generating the proper 
baryon asymmetry of the Universe.  The best such choice is identified and 
discussed.
\end{abstract}

\newpage
\baselineskip 24pt

In the Standard Model of particle interactions, the neutrino is part of a 
left-handed doublet $(\nu,l)_L$ under $SU(2)_L \times U(1)_Y$.  Whereas the 
charged lepton must have a right-handed singlet counterpart $l_R$, the 
singlet $\nu_R$ is not mandatory [because it is trivial under $SU(2)_L \times 
U(1)_Y$] and is absent in the minimal version of the model.  On the other 
hand, its existence is usually assumed so that $\nu_L$ may acquire a naturally 
small Majorana mass as $\nu_R$ gets a large Majorana mass [again because it 
is trivial under $SU(2)_L \times U(1)_Y$] in the famous canonical seesaw 
mechanism \cite{grsy,ms}.  Where does $\nu_R$ come from? and what is the 
magnitude of its Majorana mass?  The simplest answer \cite{ms} is that 
$U(1)_Y$ is actually a remnant of $SU(2)_R \times U(1)_{B-L}$ under which 
$(\nu,l)_R$ is a doublet, and the large Majorana $\nu_R$ mass comes from 
the vacuum expectation value ($vev$) of a scalar $SU(2)_R$ triplet, which also 
breaks $SU(2)_R \times U(1)_{B-L}$ to $U(1)_Y$.  This scenario has dominated 
the thinking on neutrino mass for over 20 years, but it is not the only 
possibility, even if the existence of $\nu_R$ is conceded. (Mechanisms 
without $\nu_R$ are also possible and just as natural \cite{ma98}.) It may 
not even be the best possibility as far as leptogenesis \cite{fy} is 
concerned, because the $SU(2)_R$ gauge interactions will tend to diminish 
the $\nu_R$ number density in the early Universe.

Under $SU(3)_C \times SU(2)_L \times SU(2)_R \times U(1)_{B-L}$, the quarks 
and leptons transform as: 
\begin{eqnarray}
q_L = (u,d)_L &\sim& (3,2,1,1/3), \\ 
q_R = (u,d)_R &\sim& (3,1,2,1/3), \\ 
l_L = (\nu,e)_L &\sim& (1,2,1,-1), \\ 
l_R = (\nu,e)_R &\sim& (1,1,2,-1),
\end{eqnarray}
where the electric charge is given by
\begin{equation}
Q = I_{3L} + I_{3R} + {1 \over 2} (B-L).
\end{equation}
To break $SU(2)_R \times U(1)_{B-L}$ to $U(1)_Y$, there are two possibilities. 
One is to use the scalar doublet
\begin{equation}
\Phi_R = (\phi_R^+,\phi_R^0) \sim (1,1,2,1),
\end{equation}
the other is to use the scalar triplet
\begin{equation}
\xi_R = (\xi_R^{++},\xi_R^+,\xi_R^0) \sim (1,1,3,2).
\end{equation}
The subsequent breaking of $SU(2)_L \times U(1)_Y$ to $U(1)_{em}$ may be 
achieved with either a scalar doublet
\begin{equation}
\Phi_L = (\phi_L^+,\phi_L^0) \sim (1,2,1,1),
\end{equation}
or a scalar bidoublet
\begin{equation}
\eta = \pmatrix {\eta_1^0 & \eta_2^+ \cr \eta_1^- & \eta_2^0} \sim (1,2,2,0),
\end{equation}
where $I_{3L} = 1/2, -1/2$ for the rows, and $I_{3R} = -1/2, 1/2$ for the 
columns. The existence of a scalar triplet
\begin{equation}
\xi_L = (\xi_L^{++},\xi_L^+,\xi_L^0) \sim (1,3,1,2)
\end{equation}
may also be contemplated but its $vev$ must be 
much smaller than that of $\Phi_L$ or $\eta$ to be consistent with the 
precisely determined values of $\sin^2 \theta_W$ and the masses of the 
$W$ and $Z$ bosons.  Neutrino masses are sensitive to which of these 5 
scalars are chosen, resulting in 5 basic scenarios, as described below. 

\noindent (I) $\xi_R + \eta$

This is the canonical scenario where $\nu_L$ pairs up with $\nu_R$ through 
the $vev$'s of the bidoublet $\eta$ to form a Dirac mass $m_D$ while $\nu_R$ 
picks up a large Majorana mass $m_R$ through the $vev$ of the $SU(2)_R$ 
triplet $\xi_R$.  The famous seesaw mass matrix
\begin{equation}
{\cal M}_\nu = \pmatrix {0 & m_D \cr m_D & m_R}
\end{equation}
is obtained, with $m_R$ of order the $SU(2)_R$ breaking scale.  The zero 
of this matrix comes from the fact that there is no $\xi_L$.

\noindent (II) $\xi_R + \eta + \xi_L$

This is the canonical left-right symmetric scenario, where $\xi_L 
\leftrightarrow \xi_R$ is often imposed as a symmetry of the theory.  Since 
the $vev$ of $\xi_L$ contributes to the Majorana $\nu_L$ mass, the neutrino 
mass matrix of Eq.~(11) becomes
\begin{equation}
{\cal M}_\nu = \pmatrix {m_L & m_D \cr m_D & m_R}.
\end{equation}
This means that the canonical seesaw formula is corrected to read
\begin{equation}
m_\nu = m_L - {m_D^2 \over m_R}.
\end{equation}
However $m_L$ is routinely argued to be small because $\langle \xi_L^0 
\rangle$ is of order $\langle \eta_1^0 \rangle \langle \eta_2^0 \rangle / 
\langle \xi_R^0 \rangle$ provided $m^2_{\xi_L}$ is positive and of order 
$v_R^2$.  In this case, $m_L$ may be larger or smaller than $m_D^2/m_R$, 
or the two terms may be of comparable magnitude.

For the many practitioners of the canonical seesaw mechanism, $m_L$ is 
implicitly assumed to be negligible.  On the other hand, if $m_L$ is the 
dominant term, then $\nu_R$ may be dispensed with.  In other words, we 
have just the Standard $SU(2)_L \times U(1)_Y$ Model with the simple 
addition of a Higgs triplet \cite{sv}.  Again assuming $m^2_{\xi_L}$ to be 
positive and large, we have \cite{ms98}
\begin{equation}
\langle \xi_L^0 \rangle = - \mu \langle \phi_L^0 \rangle^2 / m^2_{\xi_L},
\end{equation}
where $\mu$ is the $\xi_L^\dagger \Phi_L \Phi_L$ coupling.  This mechanism 
without any $\nu_R$ is also a 
completely satisfactory explanation of the smallness of $m_\nu$.

\noindent (III) $\Phi_R + \eta$

Here the $vev$ of $\Phi_R$ breaks $SU(2)_R \times U(1)_{B-L}$ to $U(1)_Y$, 
and all fermions obtain Dirac masses from $\eta$.  Since there is no $\xi_R$ 
or $\xi_L$, the neutrino is apparently a Dirac particle in this scenario. 
Thus $m_D$ has to be orders of magnitude smaller than any other Dirac mass.  
This is theoretically disfavored, and it is seldom discussed in the 
literature.

\noindent (IV) $\Phi_R + \eta + \Phi_L$

This is the left-right symmetric version of (III).  Again the neutrino mass 
appears to be  purely Dirac.  However, the coexistence of $\Phi_L$ and $\eta$ 
allows for an interesting extension of the usual left-right model, especially 
in the context of $E_6$.  One of the complications of using a scalar 
bidoublet in a left-right extension of the Standard Model is that two 
different $vev$'s, i.e. $\langle \eta_{1,2}^0 \rangle$, contribute to any 
given fermion mass, thus implying 
the existence of flavor changing neutral currents (FCNC) in the scalar 
sector \cite{gwp}. This is not a problem if the $SU(2)_R$ breaking scale is 
very high as in models with a large Majorana $\nu_R$ mass.  In models where 
the neutrino mass is purely Dirac, the $SU(2)_R$ breaking scale is not 
necessarily very high, so FCNC becomes the limiting constraint on the 
scale of $SU(2)_R$ breaking.  This constraint may be relaxed if there 
exists \cite{rr} an exotic quark $h$ of charge $-1/3$ such that $(u,h)_R$ 
is an $SU(2)_R$ doublet instead of the usual $(u,d)_R$.  Then 
$m_u$ comes from $\eta_1^0$, $m_d$ comes from $\phi_L^0$, and $m_h$ 
comes from $\phi_R^0$, with no FCNC in the scalar sector.  This turns out 
to be a natural possibility \cite{ma87} in the superstring-inspired $E_6$
model.  As for the lepton sector, the Dirac mass partner of $\nu_L$ is then 
a new field which is a singlet, whereas the $SU(2)_R$ partner of $e_R$ 
(usually called $\nu_R$) is now a different particle.  Because there 
are more neutral fermions in this extension, Majorana masses for $\nu_L$ 
may again be generated \cite{ma00}.

\noindent (V) $\Phi_R + \Phi_L$

This is the simplest way of breaking $SU(2)_L \times SU(2)_R \times 
U(1)_{B-L}$ to $U(1)_{em}$.  However, since the bidoublet $\eta$ is 
absent, there are apparently no fermion masses.  On the other hand, this 
creates a unique opportunity, i.e. the possibility that all fermion masses, 
be they Dirac or Majorana, come from dimension-five operators instead 
\cite{bms,fm}, i.e. operators suppressed by presumably the Planck mass. 
In the neutrino mass matrix of Eq.~(12), $m_L$ comes from $(l_L \Phi_L)^2$, 
$m_R$ comes from $(l_R \Phi_R)^2$, and $m_D$ comes from $(\bar l_L 
\Phi_L^\dagger)(l_R \Phi_R)$.  The smallness of the Majorana neutrino 
mass compared to all Dirac fermion masses may then be attributed to 
the smallness of $v_L \equiv \langle \phi_L^0 \rangle$ compared to 
$v_R \equiv \langle \phi_R^0 \rangle$.

There is another important consequence of this scenario.  Because the 
Majorana $\nu_R$ mass is now given by $v_R^2/\Lambda$, where $\Lambda$ may 
be of order the Planck mass, say $10^{19}$ GeV, it will be very much smaller 
than the $SU(2)_R$ breaking scale, i.e.
\begin{equation}
m_R \sim {v_R^2 \over \Lambda} << v_R.
\end{equation}
This means that in the early Universe, at temperatures comparable to $m_R$, 
the $SU(2)_R$ gauge interactions of $\nu_R$ are strongly suppressed and can 
safely be ignored.  This is a crucial requirement \cite{mss} for leptogenesis 
through the decay of $\nu_R$ \cite{fy}.  Recent detailed analyses \cite{lg} 
of this mechanism for obtaining a realistic baryon asymmetry of the Universe 
and its relationship to the neutrino mass matrix all assume this implicitly. 

Going back to Scenarios (III) and (IV), and allowing for the existence of 
$(l_L \Phi_L)^2$ in (IV) and that of $(l_R \Phi_R)^2$ in both (III) and 
(IV), the seesaw neutrino mass matrices of Eqs.~(11) and (12) are again 
reproduced for (III) and (IV) respectively.  This means that for a natural 
understanding of successful leptogenesis, the $SU(2)_R$ model to be adopted 
should be one with an $SU(2)_R$ doublet rather than a triplet.  Scenario (V) 
requires $v_R$ to be very high \cite{bms}, of order the grand-unification 
scale, because of $m_t$.  Scenario (IV) is a modification of (V) but 
without the $v_R$ constraint, because fermion masses may now come from $\eta$. 
Scenario (III) is a special case of (IV) and has the desirable original form 
of the seesaw neutrino mass matrix, i.e. Eq.~(11) and not Eq.~(12) as in 
(IV) and (V).  This previously neglected model should then be put forward as 
the model of choice for understanding both neutrino mass and leptogenesis.

The scalar sector of Scenario (III) consists of only $\Phi_R$ and $\eta$. 
Whereas $SU(2)_R \times U(1)_{B-L}$ is broken down to $U(1)_Y$ by the 
$vev$ of the doublet $\Phi_R$, $SU(2)_L \times U(1)_Y$ is broken down to 
$U(1)_{em}$ by the $vev$'s of the bidoublet $\eta$ which also provide Dirac 
masses for all the fermions. For example, consider the quark Yukawa couplings:
\begin{equation}
{\cal L}_Y = h_1 \bar q_L \eta q_R + h_2 \bar q_L \tilde \eta q_R + H.c.,
\end{equation}
where
\begin{equation}
\tilde \eta = \sigma_2 \eta^* \sigma_2 = \pmatrix {\bar \eta_2^0 & -\eta_1^+ 
\cr -\eta_2^- & \bar \eta_1^0}.
\end{equation}
Hence
\begin{eqnarray}
m_u &=& h_1 v_1 + h_2 v_2^*, \\ 
m_d &=& h_1 v_2 + h_2 v_1^*,
\end{eqnarray}
where $v_{1,2} \equiv \langle \eta_{1,2}^0 \rangle$.  Because
\begin{equation}
Tr \tilde \eta^\dagger \tilde \eta = Tr \eta^\dagger \eta = \bar \eta_1^0 
\eta_1^0 + \eta_1^+ \eta_1^- + \bar \eta_2^0 \eta_2^0 + \eta_2^+ \eta_2^-,
\end{equation}
whereas the independent scalar quartic terms $f_1 \Phi_R^\dagger \tilde 
\eta^\dagger \tilde \eta \Phi_R$ and $f_2 \Phi_R^\dagger \eta^\dagger 
\eta \Phi_R$ contain $f_1 v_R^2 \bar \eta_1^0 \eta_1^0$ and $f_2 v_R^2 
\bar \eta_2^0 \eta_2^0$ respectively, the effective scalar potential after 
$\Phi_R$ has been integrated out has different mass terms for $\eta_1^0$ 
and $\eta_2^0$, i.e.
\begin{equation}
(m^2 + f_1 v_R^2) \bar \eta_1^0 \eta_1^0 + (m^2 + f_2 v_R^2) \bar \eta_2^0 
\eta_2^0.
\end{equation}
This means that unless $f_1=f_2$, it is impossible to make both coefficients 
negative and of order the electroweak breaking scale.  In other words, either 
$\eta_1^0$ or $\eta_2^0$ must remain heavy, i.e. of order $v_R$.  Let 
$m^2 + f_1 v_R^2 = -\mu^2$, then $m_2^2 = (f_2-f_1)v_R^2 - \mu^2$, and 
$v_2/v_1$ is expected to be suppressed by a factor of order $v_1^2/m_2^2$, 
as shown for example in Ref.\cite{ma01}.  As a result, the contributions 
of $v_2$ to Eqs.~(18) and (19) are negligible and the suppression of FCNC 
is achieved.

The model so far has only Dirac fermion masses.  Majorana neutrino masses 
would normally come from the well-known dimension-five operator \cite{wein} 
$(l_L \Phi_L)^2$, but since $\Phi_L$ is absent, only $(l_R \Phi_R)^2$ 
is available.  Thus the original seesaw matrix of Eq.~(11) is obtained, and
$m_R$ is guaranteed to be suppressed relative to $v_R$ as shown in Eq.~(15). 
Given the particle content of Scenario (III) and the acceptance of 
higher-dimensional operators, $m_L$ actually gets a contribution from 
the dimension-seven operator $(\bar l_L \eta \tilde \Phi_R)^2/\Lambda^3$.  
However its magnitude is $v_1^2 v_R^2/\Lambda^3$ which is smaller than the 
double seesaw \cite{bms} mass of $v_1^2 \Lambda/v_R^2$ by the factor 
$(v_R/\Lambda)^4$.

Consider next the supersymmetric version of Scenario (III).  Using the 
convention that all superfields are left-handed, $q_R$ is replaced by 
$q^c \sim (3^*,1,2,-1/3)$ and $l_R$ by $l^c \sim (1,1,2,1)$.  The Higgs 
sector now consists of the superfields $\eta$, $\Phi_R$, and 
$\Phi_R^c \sim (1,1,2,-1)$.  An extra unbroken discrete $Z_2$ symmetry 
is imposed, under which quark and lepton superfields are odd, but Higgs 
superfields are even.  This serves to distinguish $l^c$ from $\Phi_R$, and 
leads to the usual $R$ parity of most supersymmetric extensions of the 
Standard Model.

To break $SU(2)_R$ at a high scale without breaking the supersymmetry, 
consider the following superpotential:
\begin{equation}
W = M \epsilon_{ij} \Phi_{Ri} \Phi^c_{Rj} + {1 \over 2 \Lambda} (\epsilon_{ij} 
\Phi_{Ri} \Phi^c_{Rj})^2.
\end{equation}
Note that an extra nonrenormalizable term has been added.  In this case, 
the scalar potential becomes
\begin{eqnarray}
V &=& \sum_j |M \epsilon_{ij} \Phi_{Ri} + {1 \over \Lambda} (\epsilon_{i'j'} 
\Phi_{Ri'} \Phi^c_{Rj'}) \epsilon_{ij} \Phi_{Ri}|^2 \nonumber \\ 
&+& \sum_i |M \epsilon_{ij} \Phi^c_{Rj} + {1 \over \Lambda} (\epsilon_{i'j'} 
\Phi_{Ri'} \Phi^c_{Rj'}) \epsilon_{ij} \Phi^c_{Rj}|^2 + V_g,
\end{eqnarray}
where $V_g$ comes from the gauge interactions of $\Phi_R$ and $\Phi_R^c$. 
Since the neutral components of $\Phi_R$ and $\Phi_R^c$ have opposite values 
of $I_{3R}$ and $B-L$, the condition $V_g = 0$ at its minimum is satisfied 
if $\langle \Phi_R \rangle = \langle \Phi_R^c \rangle$ in the above.  
A supersymmetric vacuum $(V=0)$ is thus obtained with
\begin{equation}
v_R = \langle \Phi_R \rangle = \langle \Phi^c_R \rangle = \sqrt {M \Lambda}.
\end{equation}
This shows that $M$ of Eq.~(22) may be identified with $m_R$ of Eq.~(15), i.e. 
the large Majorana mass of $\nu_R$.

In this scenario, $SU(2)_R$ is broken at the scale $v_R$ of Eq.~(24).  Below 
it, a consistent supersymmetric extension of the Standard Model survives, 
but with a singlet neutrino of Majorana mass $\sim M << v_R$.  This singlet 
neutrino couples to $(\nu \eta_2^0 - e \eta_2^+)$ but its interaction with 
the $SU(2)_R$ gauge bosons are very much suppressed at the time of the early 
Universe when its temperature is comparable to $M$.  Its decay will thus 
generate a lepton asymmetry \cite{fy,lg} which gets converted 
\cite{krs} into the present observed baryon asymmetry of the Universe 
through sphalerons during the electroweak phase transition.

So far the scale $v_R$ has not been determined.  There are two possible 
approaches.  One is to assume that it has to do with gauge-coupling 
unification of the minimal supersymmetric standard model (MSSM) \cite{review}, 
in which case it should be $10^{16}$ GeV.  This implies a singlet neutrino 
mass of order $(10^{16})^2/10^{19} = 10^{13}$ GeV.  The other is to use 
present neutrino data \cite{atm,sol,react} together with the requirement that 
the canonical seesaw matrix of Eq.~(11) yields a satisfactory baryon 
asymmetry of the Universe through $\nu_R$ decay, from which a lower bound 
on $v_R$ may be obtained.  Recent indications \cite{lg} are that the smallest 
$m_R$ is of order $10^8$ GeV, which implies that $v_R$ is at least of order 
$10^{14}$ GeV.

The bidoublet $\eta$ contains two electroweak doublets and they are just 
right for the unfication of gauge couplings in the MSSM.  However, 
two such bidoublets are usually assumed in a supersymmetric model in order 
to have realistic quark and lepton masses.  [Because of supersymmetry, we 
cannot use $\tilde \eta$ as the second bidoublet as in Eqs.~(18) and (19).] 
In that case, there are four electroweak doublets and two would have to be 
heavy (i.e. at the $v_R$ scale) not to spoil the unification of gauge 
couplings.  An alternative possibility is to keep only one bidoublet and 
invoke a flavor-nondiagonal soft supersymmetry breaking scalar sector to 
account for the observed quark and lepton mass matrices \cite{bdm}.

Scenario (III) is distinguished by the absence of a $\Phi_L$ doublet.  This 
is a problem if $SU(3)_C \times SU(2)_L \times SU(2)_R \times U(1)_{B-L}$ is 
embedded in a larger symmetry such as $SO(10)$ or $[SU(3)]^3$, because any 
scalar multiplet of this larger symmetry would also contain $\Phi_L$ if it 
contains $\Phi_R$.  In that case, Eq.~(12) is obtained, where $m_L$ comes 
from $(l_L \Phi_L)^2$.  As long as $m_L << m_D^2/m_R$, which holds if 
$\Lambda$ is of order $10^{19}$ GeV, this is an acceptable scenario as well.

In conclusion, to understand both neutrino masses in terms of the original 
canonical seesaw mechanism, i.e. Eq.~(11), and the success of leptogenesis 
through $\nu_R$ decay, the simplest and most natural model is $SU(3)_C \times 
SU(2)_L \times SU(2)_R \times U(1)_{B-L}$ with a scalar sector consisting of 
only an $SU(2)_R$ doublet $\Phi_R$ and an $SU(2)_L \times SU(2)_R$ bidoublet 
$\eta$.  The dimension-five operator $(l_R \Phi_R)^2/2\Lambda$ leads to a 
large Majorana mass for $\nu_R$ such that $m_R \simeq v_R^2/\Lambda << v_R$, 
which is desirable for generating the proper baryon asymmetry of the Universe. 
A successful supersymmetric version of this model has also been discussed.\\

This work was supported in part by the U.~S.~Department of Energy
under Grant No.~DE-FG03-94ER40837.\\

\newpage
\bibliographystyle{unsrt}

\begin{thebibliography}{99}
\bibitem{grsy} M. Gell-Mann, P. Ramond, and R. Slansky, in 
{\it Supergravity}, edited by P. van Nieuwenhuizen and D. Z. Freedman 
(North-Holland, Amsterdam, 1979), p.~315; T. Yanagida, in {\it Proceedings 
of the Workshop on the Unified Theory and the Baryon Number in the Universe}, 
edited by O. Sawada and A. Sugamoto (KEK, Tsukuba, Japan, 1979), p.~95. 
\bibitem{ms} R. N. Mohapatra and G. Senjanovic, Phys. Rev. Lett. {\bf 44}, 
912 (1980).
\bibitem{ma98} E. Ma, Phys. Rev. Lett. {\bf 81}, 1171 (1998).
\bibitem{fy} M. Fukugita and T. Yanagida, Phys. Lett. {\bf B174}, 45 (1986).
\bibitem{sv} J. Schechter and J. W. F. Valle, Phys. Rev. {\bf D22}, 2227 
(1980).
\bibitem{ms98} E. Ma and U. Sarkar, Phys. Rev. Lett. {\bf 80}, 5716 (1998).
\bibitem{gwp} S. L. Glashow and S. Weinberg, Phys. Rev. {\bf D15}, 1958 
(1977); E. A. Paschos, ibid. {\bf D15}, 1966 (1977).
\bibitem{rr} P. Ramond and D. B. Reiss, Phys. Lett. {\bf B80}, 87 (1978).
\bibitem{ma87} E. Ma, Phys. Rev. {\bf D36}, 274 (1987).
\bibitem{ma00} E. Ma, Phys. Rev. {\bf D62}, 093022 (2000).
\bibitem{bms} B. Brahmachari, E. Ma, and U. Sarkar, Phys. Rev. Lett. {\bf 91}, 
011801 (2003).
\bibitem{fm} J.-M. Frere and E. Ma, hep-ph/0305155.
\bibitem{mss} E. Ma, S. Sarkar, and U. Sarkar, Phys. Lett. {\bf B458}, 73 
(1999).
\bibitem{lg} See for example S. Davidson, hep-ph/0302075; W. Buchmuller, 
P. Di Bari, and M. Plumacher, hep-ph/0302092; E. Kh. Akhmedov, M. Frigerio, 
and A. Yu. Smirnov, hep-ph/0305322.
\bibitem{ma01} E. Ma, Phys. Rev. Lett. {\bf 86}, 2502 (2001).
\bibitem{wein} S. Weinberg, Phys. Rev. Lett. {\bf 43}, 1566 (1979).
\bibitem{krs} V. A. Kuzmin, V. A. Rubakov, and M. E. Shaposhnikov, Phys. 
Lett. {\bf B155}, 36 (1985).
\bibitem{review} See for example W. de Boer, Prog. Part. Nucl. Phys. {\bf 33}, 
201 (1994).
\bibitem{atm} C. K. Jung, C. McGrew, T. Kajita, and T. Mann, Ann. Rev. Nucl. 
Part. Sci. {\bf 51}, 451 (2001).
\bibitem{sol} Q. R. Ahmad {\it et al.}, SNO Collaboration, Phys. Rev. Lett. 
{\bf 89}, 011301, 011302 (2002); K. Eguchi {\it et al.}, KamLAND 
Collaboration, Phys. Rev. Lett. {\bf 90}, 021802 (2003).
\bibitem{react} M. Apollonio {\it et al.}, Phys. Lett. {\bf B466}, 415 (1999); 
F. Boehm {\it et al.}, Phys. Rev. {\bf D64}, 112001 (2001).
\bibitem{bdm} K. S. Babu, B. Dutta, and R. N. Mohapatra, Phys. Rev. {\bf D60}, 
095004 (1999).

\end{thebibliography}

\end{document}